\documentclass[aps,prl,showpacs,twocolumn,groupedaddress,superscriptaddress] 
{revtex4-1} 
\usepackage{graphicx} 
\usepackage{dcolumn} 
\usepackage{longtable} 
\usepackage{amsmath} 

\begin{document} 

\title{Search for the neutron-rich  hypernucleus  $^{9}_{\Lambda}$He} 

\author{M.~Agnello}\affiliation{Dipartimento di Fisica, Politecnico di Torino, 
corso Duca degli Abruzzi 24, Torino, Italy}\affiliation{INFN Sezione di 
Torino, via P. Giuria 1, Torino, Italy} \author{L.~Benussi}\affiliation 
{Laboratori Nazionali di Frascati dell'INFN, via E. Fermi 40, Frascati, Italy} 
\author{M.~Bertani}\affiliation{Laboratori Nazionali di Frascati dell'INFN, 
via E. Fermi 40, Frascati, Italy} \author{H.C.~Bhang}\affiliation{Department 
of Physics, Seoul National University, 151-742 Seoul, South Korea} 
\author{G.~Bonomi}\affiliation{Dip. di Ingegneria Meccanica e Industriale, 
Universit\`a di Brescia, via Valotti 9, Brescia, Italy} 
\affiliation{INFN Sezione di Pavia, via Bassi 6, Pavia, Italy} 
\author{E.~Botta}\thanks{Corresponding author: Elena Botta, botta@to.infn.it} 
\affiliation{Dipartimento di Fisica, Universit\`a di Torino, via P. Giuria 1, 
Torino, Italy}\affiliation{INFN Sezione di Torino, via P. Giuria 1, Torino, 
Italy} \author{M.~Bregant}\affiliation{SUBATECH, \'Ecole des Mines de Nantes, 
Universit\'e de Nantes, CNRS-IN2P3, Nantes, France} \author{T.~Bressani} 
\affiliation{Dipartimento di Fisica, Universit\`a di Torino, 
via P. Giuria 1, Torino, Italy}\affiliation{INFN Sezione di Torino, via 
P. Giuria 1, Torino, Italy} \author{S.~Bufalino}\affiliation{INFN Sezione 
di Torino, via P. Giuria 1, Torino, Italy} \author{L.~Busso}\affiliation 
{Dipartimento di Fisica, Universit\`a di Torino, via P. Giuria 1, 
Torino, Italy}\affiliation{INFN Sezione di Torino, via P. Giuria 1, Torino, 
Italy} \author{D.~Calvo}\affiliation{INFN Sezione di Torino, via P. Giuria 1, 
Torino, Italy} \author{P.~Camerini}\affiliation{INFN Sezione di Trieste, via 
Valerio 2, Trieste, Italy}\affiliation{Dipartimento di Fisica, Universit\`a 
di Trieste, via Valerio 2, Trieste, Italy} \author{B.~Dalena}\affiliation 
{CEA, Irfu/SACM, Gif-sur-Yvette, France} \author{F.~De Mori}\affiliation 
{Dipartimento di Fisica, Universit\`a di Torino, via P. Giuria 1, 
Torino, Italy}\affiliation{INFN Sezione di Torino, via P. Giuria 1, Torino, 
Italy} \author{G.~D'Erasmo}\affiliation{Dipartimento di Fisica Universit\`a 
di Bari, via Amendola 173, Bari, Italy}\affiliation{INFN Sezione di Bari, 
via Amendola 173, Bari, Italy} \author{F.L.~Fabbri}\affiliation{Laboratori 
Nazionali di Frascati dell'INFN, via E. Fermi 40, Frascati, Italy} 
\author{A.~Feliciello}\affiliation{INFN Sezione di Torino, via P. Giuria 1, 
Torino, Italy} \author{A.~Filippi}\affiliation{INFN Sezione di Torino, via 
P. Giuria 1, Torino, Italy} \author{E.M.~Fiore}\affiliation{Dipartimento di 
Fisica Universit\`a di Bari, via Amendola 173, Bari, Italy}\affiliation{INFN 
Sezione di Bari, via Amendola 173, Bari, Italy} \author{A.~Fontana} 
\affiliation{INFN Sezione di Pavia, via Bassi 6, Pavia, Italy} \author 
{H.~Fujioka}\affiliation{Department of Physics, Kyoto University, Sakyo-ku, 
Kyoto, Japan} \author{P.~Genova}\affiliation{INFN Sezione di Pavia, via Bassi 
6, Pavia, Italy} \author{P.~Gianotti}\affiliation{Laboratori Nazionali di 
Frascati dell'INFN, via E. Fermi 40, Frascati, Italy} \author{N.~Grion} 
\affiliation{INFN Sezione di Trieste, via Valerio 2, Trieste, Italy} 
\author{V.~Lucherini}\affiliation{Laboratori Nazionali di Frascati dell'INFN, 
via E. Fermi 40, Frascati, Italy} \author{S.~Marcello}\affiliation 
{Dipartimento di Fisica, Universit\`a di Torino, via P. Giuria 1, 
Torino, Italy}\affiliation{INFN Sezione di Torino, via P. Giuria 1, Torino, 
Italy} \author{N.~Mirfakhrai}\affiliation{Department of Physics, Shahid 
Behesty University, 19834 Teheran, Iran} 
\author{F.~Moia}\affiliation{Dip. di Ingegneria Meccanica e Industriale, 
Universit\`a di Brescia, via Valotti 9, Brescia, Italy} 
\affiliation{INFN Sezione di Pavia, via Bassi 6, Pavia, Italy}
\author{O.~Morra}\affiliation{INAF-IFSI, Sezione di Torino, Corso Fiume 4, 
Torino, Italy}\affiliation{INFN Sezione di Torino, via P. Giuria 1, Torino, 
Italy} \author{T.~Nagae}\affiliation{Department of Physics, Kyoto University, 
Sakyo-ku, Kyoto, Japan} \author{H.~Outa}\affiliation{RIKEN, Wako, Saitama 
351-0198, Japan} \author{A.~Pantaleo}\thanks{deceased}\affiliation{INFN 
Sezione di Bari, via Amendola 173, Bari, Italy} \author{V.~Paticchio}
\affiliation{INFN Sezione di Bari, via Amendola 173, Bari, Italy} 
\author{S.~Piano}\affiliation{INFN Sezione di Trieste, via Valerio 2, Trieste, 
Italy} \author{R.~Rui}\affiliation{INFN Sezione di Trieste, via Valerio 2, 
Trieste, Italy}\affiliation{Dipartimento di Fisica, Universit\`a di Trieste, 
via Valerio 2, Trieste, Italy} \author{G.~Simonetti}\affiliation{Dipartimento 
di Fisica Universit\`a di Bari, via Amendola 173, Bari, Italy}\affiliation 
{INFN Sezione di Bari, via Amendola 173, Bari, Italy} \author{R.~Wheadon} 
\affiliation{INFN Sezione di Torino, via P. Giuria 1, Torino, Italy} 
\author{A.~Zenoni}\affiliation{Dip. di Ingegneria Meccanica e Industriale, 
Universit\`a di Brescia, via Valotti 9, Brescia, Italy} 
\affiliation{INFN Sezione di Pavia, via Bassi 6, Pavia, Italy} 
\collaboration{The FINUDA Collaboration} 
\noaffiliation 
\author {A.~Gal} 
\affiliation{Racah Institute of Physics, The Hebrew University, Jerusalem 
91904, Israel} 

\date{\today}      
\begin{abstract} 
Search for the neutron-rich hypernucleus $_{\Lambda}^{9}{\rm He}$ is 
reported by the FINUDA experiment at DA$\Phi$NE, INFN-LNF, studying 
$(\pi^+,\pi^-)$ pairs in coincidence from the $K^{-}_{\rm stop}+{^{9}{\rm Be}} 
\rightarrow {^{9}_{\Lambda}{\rm He}}+\pi^{+}$ production reaction followed by 
$_{\Lambda}^{9}{\rm He}\to {^{9}{\rm Li}}+\pi^-$ weak decay. An upper limit 
of the production rate of $_{\Lambda}^{9}{\rm He}$ undergoing this two-body 
$\pi^-$ decay is determined to be $(2.3\pm 1.9)\cdot 10^{-6}/K^{-}_{\rm stop}$ 
at 90$\%$ confidence level. 
\end{abstract} 

%Hypernuclei, Kaon-induced reactions, Nucleon distributions and halo features 

\maketitle 

In recent papers we reported \cite{PRLnrich} and described in detail 
\cite{NPAnrich} the first experimental evidence for the existence of hyper 
superheavy hydrogen $^{6}_{\Lambda}$H. Three candidate events for such 
a particle-stable nuclear system were uniquely identified in the FINUDA 
experiment at DA$\Phi$NE, Frascati (Italy) by observing $\pi^{+}$ mesons 
from the ($K^{-},\pi^{+}$) reaction on $^{6}$Li targets, in coincidence with 
$\pi^{-}$ mesons from $^{6}_{\Lambda}{\rm H}\rightarrow{^{6}{\rm He}}+\pi^{-}$ 
weak decay. The $^{6}_{\Lambda}$H binding energy with respect to $^{5}{\rm H}+
\Lambda$ was determined jointly from production and decay processes to be 
$B_{\Lambda}({^{6}_{\Lambda}{\rm H}})=(4.0\pm 1.1)$ MeV, assuming that 
the $^{5}$H ground-state (g.s.) resonance lies at 1.7 MeV above the 
$^{3}{\rm H}+2n$ lowest neutron emission threshold \cite{5H}. We remark that 
$^{6}_{\Lambda}$H is a particle-stable nuclear system with the highest $N/Z=4$ 
value ($(N+\Lambda)/Z=5$) measured so far, higher than for the archetype 
neutron-rich nucleus $^{11}$Li. Since $^{7}$Li and $^{9}$Be targets were used 
in the same data taking in which $^{6}_{\Lambda}$H was produced on $^{6}$Li 
targets, with a similar number of stopped $K^{-}$, we examined whether the 
method applied to the successful search for $^{6}_{\Lambda}$H could be 
extended to $^{7}_{\Lambda}$H and $^{9}_{\Lambda}$He. The case of 
$^{7}_{\Lambda}$H was dismissed, since the daughter nucleus $^{7}$He produced 
in the two-body weak decay $^{7}_{\Lambda}{\rm H}\rightarrow {^{7}{\rm He}}+
\pi^{-}$ is particle-unstable, making non-applicable the experimental method 
that will be briefly outlined in the following. However, the method could be 
applied in the case of $^{9}_{\Lambda}$He, since both $^{9}$Li g.s. and first 
excited state at 2.691 MeV are particle stable \cite{lifetime}, allowing thus 
a two-body weak decay $^{9}_{\Lambda}{\rm He}\to {^{9}{\rm Li}}+\pi^{-}$. 

\begin{figure}[h] 
\vspace{-4mm} 
\begin{center} 
\includegraphics[width=0.45\textwidth]{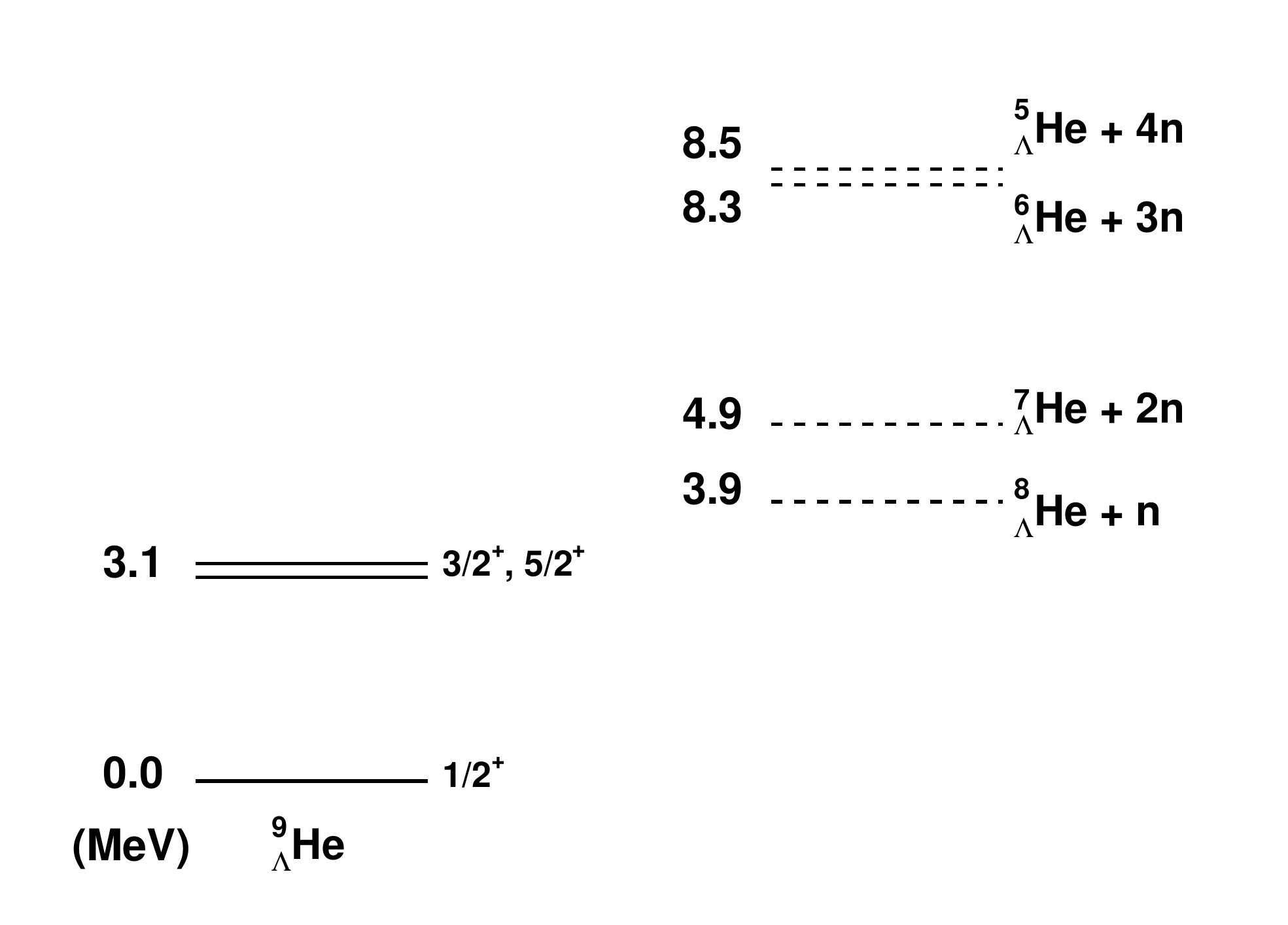} 
\caption{Anticipated $^{9}_{\Lambda}$He energy level scheme below the lowest 
neutron emission threshold, together with higher neutron emission thresholds. 
Note the schematically marked $^{9}_{\Lambda}$He excited doublet which is 
based on $^{8}$He (particle-unstable) first excitation $2^+$ at $\approx 3.1$ 
MeV \cite{lifetime}.} 
\label{fig1} 
\end{center} 
\vspace{-3mm} 
\end{figure}

The neutron-rich $^{9}_{\Lambda}$He hypernucleus is one of the exotic 
$\Lambda$-hypernuclear species considered decades ago by Dalitz and Levi Setti 
\cite{dalitz}, and by Majling \cite{majling} who estimated the binding-energy 
$B_{\Lambda}({^{9}_{\Lambda}{\rm He}})=8.5$ MeV. This value, coinciding 
with $B_{\Lambda}({^{9}_{\Lambda}{\rm Li}})$ \cite{davis}, is based on the 
assumption that the increased neutron excess in $^{9}_{\Lambda}$He with 
respect to $^{9}_{\Lambda}$Li does not induce irregularities in the known 
binding energy systematics. The assumption is consistent with the similarity 
of $B_{\Lambda}$ values for $^{6}_{\Lambda}$H \cite{PRLnrich,NPAnrich} and 
$^{6}_{\Lambda}$He \cite{davis}, and for $^{7}_{\Lambda}$He \cite{hashimoto} 
and $^{7}_{\Lambda}$Li \cite{davis}. Millener's recent shell-model study of 
$p$-shell $B_{\Lambda}$ values, Table 2 in Ref.~\cite{millener}, suggests that 
$B_{\Lambda}({^{9}_{\Lambda}{\rm He}})\approx B_{\Lambda}({^{9}_{\Lambda}{\rm 
Li}})$ to within less than 0.1 MeV. We shall therefore adopt the value 
$B_{\Lambda}({^{9}_{\Lambda}{\rm He}})=8.5$ MeV as a working hypothesis. 
Fig.~\ref{fig1} shows the expected particle-stable $^{9}_{\Lambda}$He levels, 
together with the neutron emission thresholds below the 8.5 MeV $\Lambda$ 
emission threshold. 

We outline now briefly the experimental method adopted in the search 
for $^{9}_{\Lambda}$He. The DA$\Phi$NE $\phi$-Factory in Frascati uses 
$e^{+}e^{-}$ collisions at total c.m. energy $\sqrt{s}=1020$ MeV to produce 
$\phi$ mesons that decay into ($K^{+},K^{-}$) pairs with a $49\%$ branching 
ratio. The resulting $K^{-}$ mesons of kinetic energy $16.1\pm 0.5$ MeV 
can be stopped in nuclear targets. In the FINUDA experiment, a total of 
$2.5 \cdot 10^{7}$ $K^{-}$ mesons were detected as stopped in two $^{9}$Be 
targets, 2 mm thick. The FINUDA detector has been described in detail recently 
in \cite{NPAnrich,NMWDp}. $^{9}_{\Lambda}$He can be produced in the two-body 
reaction: 
\begin{equation} 
K^{-}_{\rm stop}+{^{9}{\rm Be}}\rightarrow {^{9}_{\Lambda}{\rm He}}+\pi^{+}\;. 
\label{prod} 
\end{equation} 
Assuming $B_{\Lambda}({^{9}_{\Lambda}{\rm He}})=8.5$ MeV, it is 
straightforward to evaluate the momentum $p_{\pi^{+}}=257.5$ MeV/c and 
kinetic energy $T_{\pi^{+}}=153.3$ MeV for a $\pi^{+}$ meson emitted in 
(\ref{prod}). The two-body weak decay 
\begin{equation} 
{^{9}_{\Lambda}{\rm He}_{\rm g.s.}}\rightarrow {^{9}{\rm Li}_{\rm g.s.}}+
\pi^{-} 
\label{decay} 
\end{equation} 
should then produce a $\pi^{-}$ meson with $p_{\pi^{-}}=116.9$ MeV/c and 
$T_{\pi^{-}}=42.5$ MeV. We note that $^{9}_{\Lambda}$He could be produced 
in the reaction (\ref{prod}) also in one of the excited doublet levels 
marked schematically in Fig.~\ref{fig1} that, if particle-stable, would 
$\gamma$-decay to $^{9}_{\Lambda}{\rm He}_{\rm g.s.}$ which then decays 
weakly according to (\ref{decay}). However, one or both of these 
$^{9}_{\Lambda}$He doublet levels could prove to be isomeric, similar to what 
is believed to occur for $^{7}_{\Lambda}$He \cite{davis}. One has to allow 
for such a scenario when considering the spread of the $\pi^{\pm}$ accepted 
momenta and kinetic energies. 

The formation (\ref{prod}) and decay (\ref{decay}) reactions occur both at 
rest, since the stopping time of $^{9}_{\Lambda}$He in the material (Be) is 
shorter than its lifetime which is of the order of 2.6$\cdot 10^{-10}$ s (the free $\Lambda$ lifetime). 
Momentum conservation is then automatically ensured and energy conservation 
is expressed explicitly for (\ref{prod}): 
\begin{eqnarray} 
M(K^{-})+4M(p)+5M(n)-B(^{9}{\rm Be}) & =&   \nonumber \\ 
M({^{9}_{\Lambda}{\rm He}})+
T({^{9}_{\Lambda}{\rm He}})+M(\pi^{+})+T(\pi^{+}) \;, 
\label{Eform} 
\end{eqnarray} 
and for (\ref{decay}): 
\begin{eqnarray} 
M({^{9}_{\Lambda}{\rm He}})=3M(p)+6M(n)-&B({^{9}{\rm Li}})+T({^{9}{\rm Li}})& 
\nonumber  \\ 
+M(\pi^{-})+T(\pi^{-}) \;, 
\label{Edecay} 
\end{eqnarray} 
in which $M$ stands for mass, $T$ for kinetic energy, and $B$ for nuclear 
binding energy. Combining Eqs.~(\ref{Eform}) and (\ref{Edecay}) in order to 
eliminate $M({^{9}_{\Lambda}{\rm He}})$, we get the following equation: 
\begin{eqnarray} 
T(\pi^{+}) & + &T(\pi^{-})  =   M(K^{-}) + M(p) - M(n) - 2M(\pi) \nonumber \\ 
& & - B({^{9}{\rm Be}}) + B({^{9}{\rm Li}}) -T({^{9}{\rm Li}})-
T({^{9}_{\Lambda}{\rm He}}) \;. 
\label{Ebal} 
\end{eqnarray} 
All the terms on the right hand side are known constants except for 
$T({^{9}_{\Lambda}{\rm He}})$ and $T({^{9}{\rm Li}})$ that can be evaluated 
from momentum and energy conservation and depend on the unknown value of 
$B_{\Lambda}({^{9}_{\Lambda}{\rm He}})$. 

A variation of $B_{\Lambda}({^{9}_{\Lambda}{\rm He}})$ between 0 and 10 MeV 
introduces a change of $\sim$ 0.1 MeV in $T(\pi^{+})+T(\pi^{-})$ (\ref{Ebal}), 
corresponding to a sensitivity of 10 keV per MeV of 
$B_{\Lambda}({^{9}_{\Lambda}{\rm He}})$. 
This change is much smaller than the measured energy resolutions for 
$\pi^{+}$ (deduced from the 235.6 MeV/c monochromatic $\mu^{+}$ line in 
$K_{\mu2}$ decay) and $\pi^{-}$ (deduced from the 132.8 MeV/c monochromatic 
$\pi^{-}$ line in the two-body $^{4}_{\Lambda}$H mesonic decay): 
$\sigma_{T(\pi^{+})}=0.96$ MeV and $\sigma_{T(\pi^{-})}=0.84$ MeV. 
The FINUDA energy resolution for a ($\pi^{+},\pi^{-}$) pair in coincidence 
is $\sigma_{T}=1.3$ MeV \cite{NPAnrich}. We assume a value of 
$B_{\Lambda}({^{9}_{\Lambda}{\rm He}})=8.5$ MeV \cite{majling}; 
therefore $T_{\rm sum}\equiv T(\pi^{+})+T(\pi^{-})=195.8\pm 1.3$ MeV. 

\begin{figure}[h] 
\vspace{-2mm} 
\begin{center} 
%\hspace{-2mm}
%\includegraphics[width=0.45\textwidth]{Ppim_Ppip_cut_9Be2.pdf} 
\includegraphics[width=0.53\textwidth]{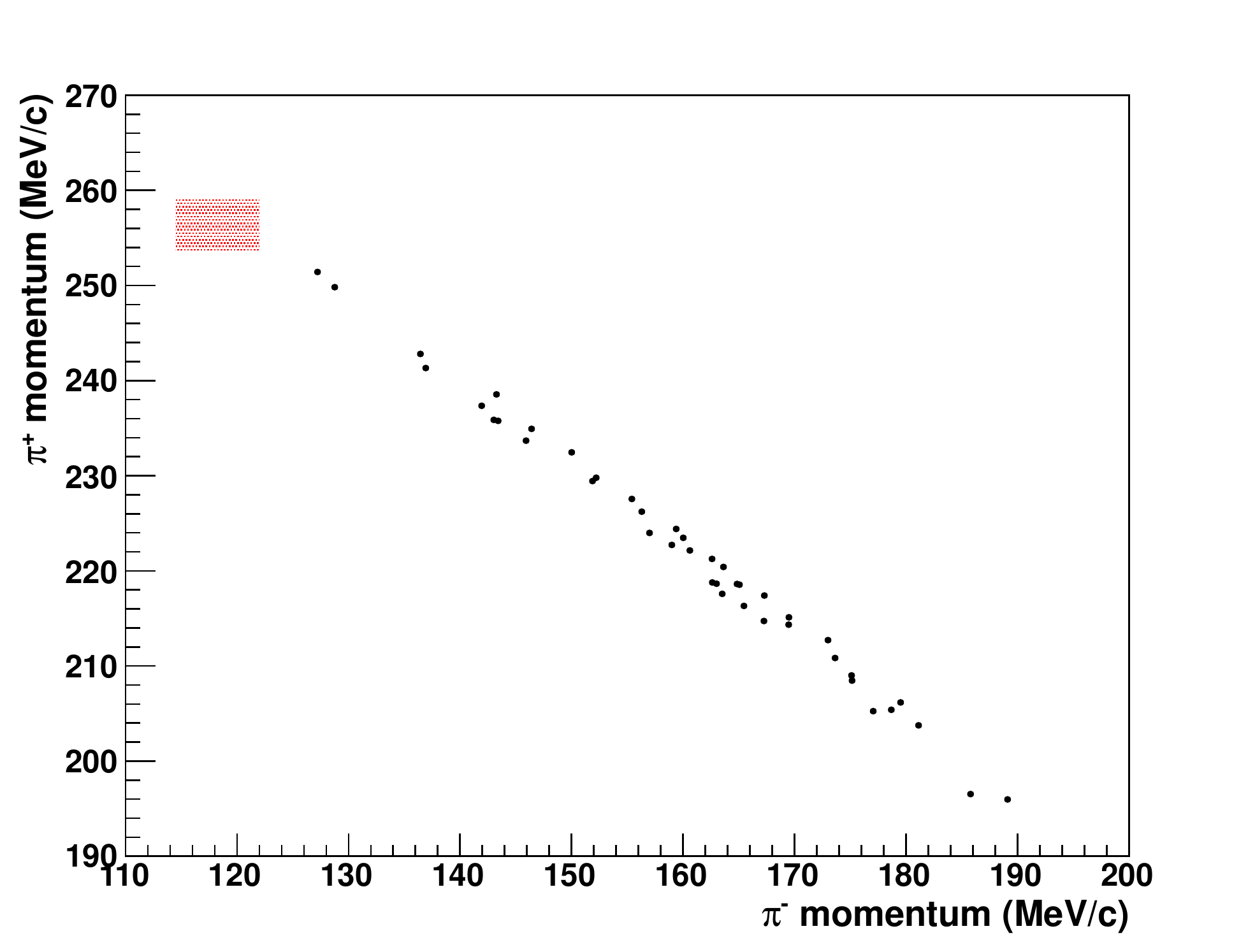} 
\caption{(color online). $\pi^{+}$ momentum vs $\pi^{-}$ momentum for 
$^{9}$Be target events with $T_{\rm sum}=(194.5-197.5)$ MeV. The shaded 
(red) rectangle indicates the position of events with $p_{\pi^{+}}=
(253.5-259)$ MeV/c and $p_{\pi^{-}}=(114.5-122)$ MeV/c.} 
\label{fig2} 
\end{center} 
\vspace{-6mm} 
\end{figure} 

Then we consider, for the coincidence ($\pi^{+},\pi^{-}$) events, only those 
for which the sum of the kinetic energies $T_{\rm sum}$ assumes values in the 
range ($194.5-197.5$) MeV. The half-width of this interval 
corresponds to 1.15 $\sigma_{T}$, in order to be selective on possible 
background events and benefiting from the excellent stability of the FINUDA 
magnetic spectrometry. A two-dimensional plot of these selected events 
is shown in Fig.~\ref{fig2}. Events associated with the formation of 
$^{9}_{\Lambda}$He should fall in the hatched (red) rectangle in the figure, 
with $p_{\pi^{+}}=(253.5-259)$ MeV/c and $p_{\pi^{-}}=(114.5-122)$ MeV/c. 
These values correspond to pion momenta that span values of 
$B_{\Lambda}({^{9}_{\Lambda}{\rm He}})$ between 5 and 10 MeV. 

\begin{figure*}[ht] 
%\vspace{-5mm} 
\begin{center} 
\includegraphics[width=0.45\textwidth]{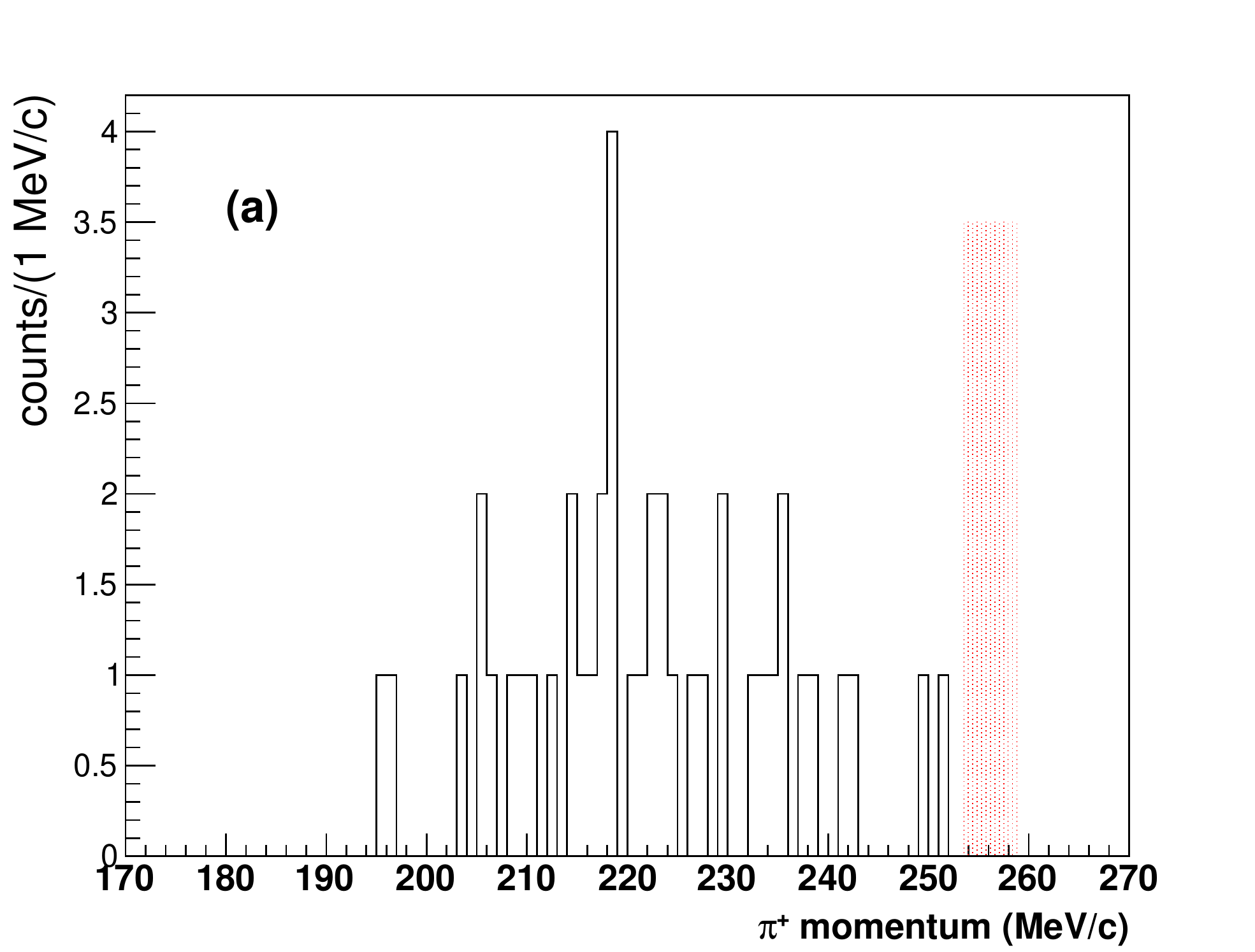} 
\hspace{2mm} 
\includegraphics[width=0.45\textwidth]{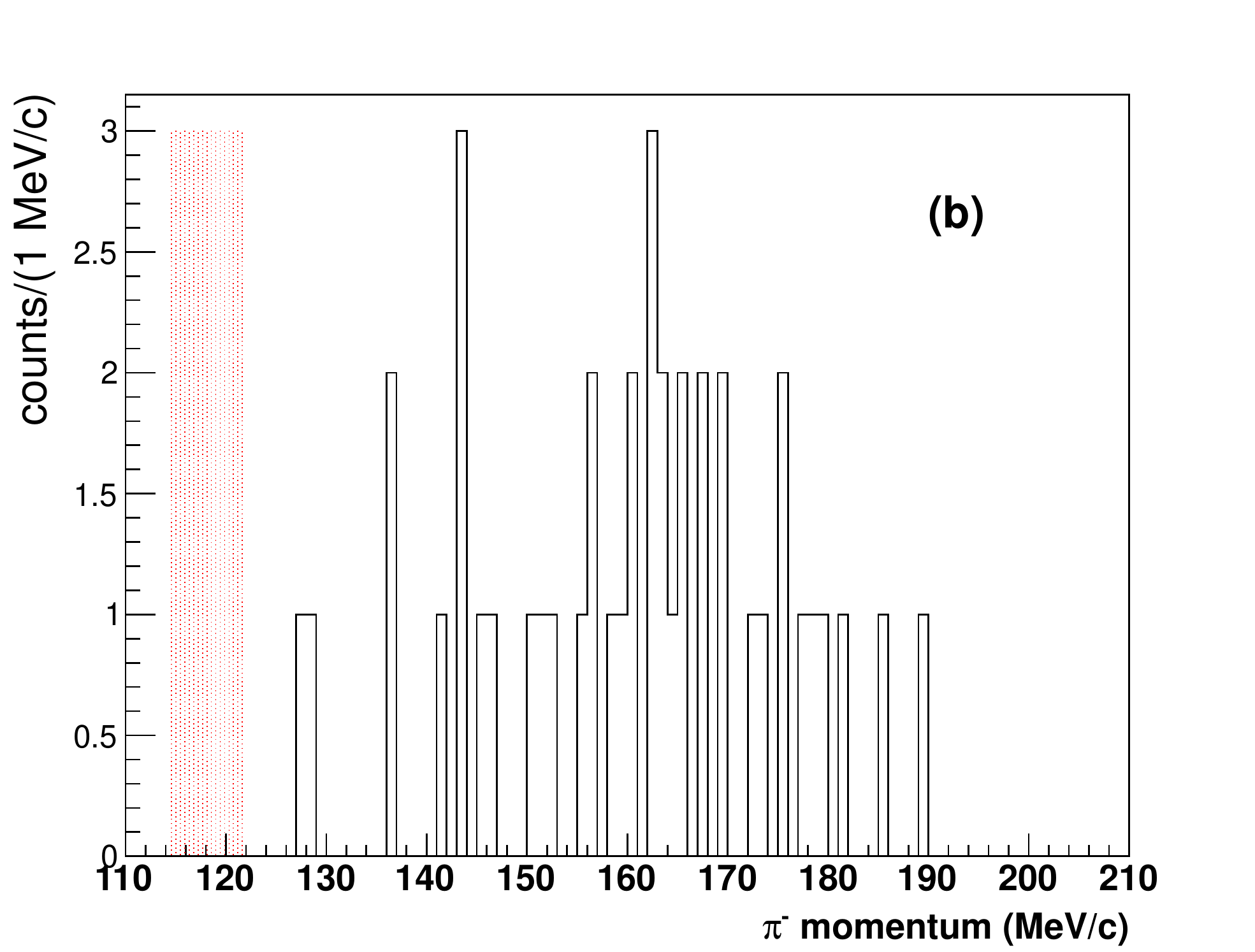} 
\caption{(color online). $\pi^{+}$ momentum (a) and $\pi^{-}$ momentum (b) 
distributions for $^{9}$Be target events with $T_{\rm sum}=(194.5-197.5)$ MeV. 
The shaded (red) rectangles highlight pion momenta $p_{\pi^{+}}=(253.5-259)$ 
MeV/c and $p_{\pi^{-}}=(114.5-122)$ MeV/c, corresponding to 
$B_{\Lambda}({^{9}_{\Lambda}{\rm He}})=5-10$ MeV.} 
\label{fig3} 
\end{center} 
%\vspace{-5mm} 
\end{figure*} 

Fig.~\ref{fig3} shows the projections on the two axes of the distribution 
of Fig.~\ref{fig2}: there are clearly no events satisfying the conditions 
required by the formation and decay of 
$^{9}_{\Lambda}$He with $B_{\Lambda}({^{9}_{\Lambda}{\rm He}})\geq 5$ MeV; 
recall from Fig.~\ref{fig1} that $B_{\Lambda}({^{9}_{\Lambda}{\rm He}})$ = 5 MeV is about 1 MeV above 
the lowest neutron emission threshold expected for $^{9}_{\Lambda}$He. 

Since no events that could be attributed to the existence of a bound 
$^{9}_{\Lambda}$He were found, we did not follow the analysis done for 
$^{6}_{\Lambda}$H \cite{NPAnrich} in which extensive calculations were 
performed on possible backgrounds that would mimic the expected true events. 
Events filling the full distribution of Fig.~\ref{fig2} are certainly 
attributable primarily to reactions with the production of a quasi-free 
$\Sigma^{+}$, but it was outside the scope of the present analysis to 
reproduce the shape and strength of this distribution. 

Given the experimental procedure described above, it was possible to derive an 
upper limit for $R \cdot {\rm BR}(\pi^{-})$, where $R$ is $^{9}_{\Lambda}$He 
production rate per stopped $K^{-}$ in reaction (\ref{prod}) and BR$(\pi^{-})$ 
is the branching ratio (BR) for $^{9}_{\Lambda}$He two-body weak decay 
(\ref{decay}): 
\begin{eqnarray} 
R \cdot {\rm BR}(\pi^{-}) & \leq & \frac{N} 
{\epsilon(\pi^{+})\ \epsilon(\pi^{-})\ K^{-}_{\rm stop}({^{9}{\rm Be}})} 
\nonumber \\ 
& = & (1.6 \pm 1.3) \cdot 10^{-6} /K^{-}_{\rm stop} 
\label{rate} 
\end{eqnarray} 
Here, $N$ is the expected mean value of the observation for which a null 
observation is $10\%$ probable [upper limit at $90\%$ confidence level 
(C.L.)], $\epsilon(\pi^{+})$ and $\epsilon(\pi^{-})$ indicate the global 
efficiencies for $\pi^{+}$ and $\pi^{-}$, respectively, including detection 
efficiency, geometrical and trigger acceptances and pattern recognition, 
reconstruction and selection efficiencies, all of which have been evaluated 
by means of the full FINUDA simulation code, well tested in calculations for 
other reactions in similar momentum ranges \cite{spectrFND,NPA775,mwd}. 
$K^{-}_{\rm stop}({^{9}{\rm Be}})$ is the number of $K^{-}$ detected at stop 
in $^{9}$Be targets. 

For the evaluation of the upper limit a correction for the 1.15 $\sigma_{T}$ 
cut applied to $T(\pi^{+})+T(\pi^{-})$ has to be taken into account and 
a correction for the fraction of $^{9}_{\Lambda}$He decaying in flight 
has to be applied too, which is estimated to be smaller than 8$\%$ \cite{tamura89}. 
The $R \cdot {\rm BR}(\pi^{-})$ value, corrected for both effects is: 
$R \cdot {\rm BR}(\pi^{-})<(2.3 \pm 1.9) \cdot 10^{-6} /K^{-}_{\rm stop}$. 

To derive the upper limit $R$ for the production rate of $^{9}_{\Lambda}$He 
particle-stable levels, we need to know the branching ratio BR($\pi^{-}$) 
for the two-body weak decay $^{9}_{\Lambda}{\rm He}_{\rm g.s.} \rightarrow 
{^{9}{\rm Li}_{\rm g.s.}} + \pi^{-}$. The other possible two-body decay, 
to $^{9}$Li(2.691 MeV) with $p_{\pi^{-}}=112.6$ MeV/c corresponding to the 
value $B_{\Lambda}=8.5$ MeV assumed here, is outside of the $p_{\pi^{-}}$ 
cut imposed in the present search and, therefore, it does not contribute 
to BR($\pi^{-}$). Nevertheless, inspection of the $\pi^{-}$ momentum 
distribution in Fig.~\ref{fig3} (b) suggests that this two-body decay too 
is not observed in our measurement. In absence of published evaluations of 
the branching ratio for the weak decay $^{9}_{\Lambda}{\rm He}_{\rm g.s.} 
\rightarrow 
{^{9}{\rm Li}_{\rm g.s.}} + \pi^{-}$ in which a $1s$ $\Lambda$ 
is transformed to a $1p$ proton, we follow Ref.~\cite{gal09} and evaluate 
\begin{equation} 
\Gamma({^{9}_{\Lambda}{\rm He}_{\rm g.s.}}\to {^{9}{\rm Li}_{\rm g.s.}}+
\pi^{-})=0.261~\Gamma_{\Lambda}, 
\label{g.s.rate} 
\end{equation} 
where $\Gamma_{\Lambda}$ is the free-$\Lambda$ decay rate which approximates 
fairly the total $\Lambda$-hypernuclear decay rate in the relevant mass 
range \cite{bbg12}. For completeness, we give also the rate evaluated for 
decay to $^{9}$Li(2.691 MeV): 
\begin{equation} 
\Gamma({^{9}_{\Lambda}{\rm He}_{\rm g.s.}}\to {^{9}{\rm Li}_{\rm exc.}}+
\pi^{-})=0.094~\Gamma_{\Lambda}. 
\label{exc.rate} 
\end{equation} 
Using the branching ratio value BR$(\pi^{-})=0.261$ from (\ref{g.s.rate}), 
we obtain the following upper limit for the production of $^{9}$He: 
\begin{equation} 
R<(2.3+1.9)/0.261\cdot 10^{-6}/K^{-}_{\rm stop}=1.6 \cdot 
10^{-5}/K^{-}_{\rm stop} 
\label{R} 
\end{equation} 
at $90\%$ C.L. This improves by over an order of magnitude the previous upper 
limit set in set in an experiment performed at the High Energy Accelerator Research 
Organization, Tsukuba, Japan (KEK)  \cite{KEK96}. 

\begin{table}[h] 
\begin{center} 
\caption{Upper limits on rates $R$ per stopped $K^{-}$, 
for production of $p$-shell neutron-rich hypernuclei in the 
$(K^{-}_{\rm stop},\pi^{+})$ reaction.} 
\label{tab1} 
\begin{tabular}{cccc} 
\hline \hline 
\noalign{\smallskip} 
$^{7}_{\Lambda}$H \cite{FIN06} \; & \; $^{9}_{\Lambda}$He \cite{present} \; & 
\; $^{12}_{~\Lambda}$Be \cite{FIN06} \;&\; $^{16}_{~\Lambda}$C \cite{KEK96} \\ 
\noalign{\smallskip} 
\hline 
\noalign{\smallskip} 
$5.4 \cdot 10^{-5}$ \; & \; $1.6 \cdot 10^{-5}$ \; & \; $2.4 \cdot 10^{-5}$ \; 
& \; $6.2 \cdot 10^{-5}$ \\ 
\hline \hline 
\end{tabular} 
\end{center} 
\end{table} 

Table~\ref{tab1} summarizes the lowest upper limits on production rates 
for neutron-rich hypernuclei in the $p$ shell from searches done at 
KEK \cite{KEK96} and during the first data taking of FINUDA \cite{FIN06}; 
to compare directly to (\ref{R}), the statistical error in \cite{FIN06} has 
been added. 
These upper limits were deduced through the analysis of inclusive spectra 
of $\pi^{+}$ mesons emitted following the capture of $K^{-}$ mesons at rest 
by nuclei and looking for peaks in relevant momentum regions. We note that all 
of these upper limits do not go below the value $R=10^{-5}/K^{-}_{\rm stop}$, 
higher than the $^{6}_{\Lambda}$H production rate deduced from the 
observation of three $^{6}_{\Lambda}$H candidate events \cite{PRLnrich}. 
Clearly, the observation of neutron-rich hypernuclei by studying inclusive 
spectra of $\pi^{+}$ mesons was hindered in these experiments by the 
overwhelming background from reactions leading to the production of a 
$\Sigma^{+}$ hyperon on one or two correlated protons. The technique of 
taking in coincidence also a $\pi^{-}$ meson from the weak decay of the 
produced neutron-rich hypernucleus, while applying a narrow selection on 
the sum of the kinetic energies of the ($\pi^{+},\pi^{-}$) pair, allowed to 
distinguish for the first time $^{6}_{\Lambda}$H and to improve by over one 
order of magnitude the upper limit on the production of $^{9}_{\Lambda}$He 
reported from KEK \cite{KEK96}. We note that the method of enforcing a 
$\pi^{-}$ weak decay coincidence suffers from the theoretical uncertainty 
associated with deducing the particular two-body weak decay branching ratio 
BR. However, it was shown \cite{mwd} that the available relevant theoretical 
calculations \cite{gal09,motoba88,motoba94} are fully reliable. 

There are no detailed theoretical calculations on the production rates for 
light neutron-rich hypernuclei in $K^{-}$ capture at rest. From experiment, 
the production rate of $^{6}_{\Lambda}$H \cite{PRLnrich,NPAnrich} is two to 
three orders of magnitude lower than those for the production of bound states 
of `ordinary' light hypernuclei in ($K^{-}_{\rm stop},\pi^{-}$) reactions 
\cite{spectrFND}. These new measurements by FINUDA provide a new impetus that 
should stimulate further efforts in this field both experimentally and 
theoretically.

\end{document}